# The whereabouts of 2D gels in quantitative proteomics

Thierry Rabilloud

CEA-DSV-iRTSV/LCBM and UMR CNRS-UJF 5249

CEA Grenoble

17 rue des martyrs, F-38054 Grenoble Cedex 9, France

thierry.rabilloud@cea.fr

Abstract:

Two dimensional gel electrophoresis has been instrumental in the development of proteomics. Although it is no longer the exclusive scheme used for proteomics, its unique features make it a still highly valuable tool, especially when multiple quantitative comparisons of samples must be made, and even for large samples series. However, quantitative proteomics using two-dimensional gels is critically dependent on the performances of the protein detection methods used after the electrophoretic separations. This chapter therefore examines critically the various detection methods, (radioactivity, dyes, fluorescence and silver) as well as the data analysis issues that must be taken into account when quantitative comparative analysis of two-dimensional gels is performed.

**ii. Key Words:** 2D-PAGE, fluorescence dyes, image analysis, organic dyes, polyacrylamide gels, quantification, radioisotopes, silver staining

# 1. Introduction

Since its introduction in the mid-seventies *(1, 2)*, 2D electrophoresis has always been used as a quantitative technique of protein analysis, and it is fair to say that such quantitative analyses (e.g. in *(3-8)*) have preceded the real onset of proteomics, hallmarked by the first protein identification techniques, at that time based on Edman sequencing *(9-13)*. In the current proteomics landscape, completely dominated by tandem mass spectrometry *(14, 15)*, 2D gel-based proteomics represents an exception in the sense that this is the only proteomics setup where

i) protein quantification is not made in a mass spectrometer (and there has been only very limited attempts to break this rule *(16)*).

ii) the on-gel quantification step is often used as a screening process to select a limited set of protein spots that are then further characterized by mass spectrometry.

These two cardinal features have put an enormous pressure on the performances that the on-gel protein detection methods must show, as it is quite clear that what is not detected is never analyzed and thus completely ignored. Thus, the on-gel protein detection methods must be very sensitive, but also linear in response (to be able to detect accurately abundance variations), homogeneous (so that all classes of proteins are detected) and of course fully compatible with mass spectrometry (to ensure easy and accurate protein characterization).

Although these constraints have been quite clear to the community for numerous years, and have led to the implementation of many protein detection schemes (see below and in other chapters of this book), an often overlooked problem is the performance of the 2D electrophoresis itself. In other words, what is the quantitative yield of 2D electrophoresis and how homogeneous this yield is for various classes of proteins. There are very few papers dealing with this issue, but one recent paper *(17)* showed that the yield of 2D electrophoresis was rather moderate (20-40%), which is an often overlooked parameter when the overall efficiency of the system is considered. Furthermore, work on membrane proteins *(18, 19)* has strongly suggested that protein losses are not homogeneous and may be much greater for poorly soluble proteins such as membrane proteins.

Of course, the overall yield of the process will also depend on the efficiency of the protein extraction during the sample preparation process, but this process is so variable from one sample to another that it is really beyond the scope of this chapter. Furthermore, this chapter will deal mainly with the quantification issues in 2D gel-based proteomics. Other important issues, such as the scope of gel-based proteomics the interest of protein electrophoresis in modern proteomics and how electrophoresis systems can be modulated to improve their performances, have been reviewed elsewhere *(20-22)*.

## 2. The protein detection methods

Over the years, numerous on-gel protein detection methods have been used, each having its advantages and drawbacks in the sensitivity/linearity/homogeneity/compatibility multiple criteria selection guide. Over the numerous years of use of 2D electrophoresis, some techniques have almost disappeared, while others are currently flourishing. Many different protein detection schemes have been devised over the several decades of use of 2D gels *(23)*, and only the most important ones will be reviewed in this chapter.

### 2.1. Protein detection via radioisotopes

This is an example of a technique that has subsided now, although it played a key role in the early days of proteomics, before the name was even coined.

Due to its exquisite sensitivity *(24, 25)* and linearity *(26)*, protein detection via radioisotopes has been associated with almost all the early success stories of 2D electrophoresis, from the determination of protein numbers in cells *(27)*, to the first identification of a protein from 2D gels, i.e. proliferating cell nuclear antigen (PCNA) *(28, 29)*, to single cell proteomics *(30)* or to phosphoprotein studies *(31)*. These exquisite sensitivity and linearity have even increased with new detection technologies such as phosphor screens *(32)*. However, except for special purposes where it is almost irreplaceable *(17, 33)*, detection via radioisotopes has almost disappeared from modern proteomics. In addition to the fact that not all samples are easily accessible to this type of detection (e.g. human samples), increasing safety and regulatory issues, hastened the decline of radioactivity in proteomics.

## 2.2. Protein detection via organic dyes

In this mode of detection, the name of the game is to bind as many dye molecules per protein molecules as possible, in order to create a light absorption signal that is detectable. Of course, this process must be as reproducible as possible, and the molar extinction coefficient of the dye also plays a major role in the signal intensity.

For all these reasons, and despite a few attempts to use other dyes *(34)*, colloidal Coomassie Blue, as introduced in 1988 by Neuhoff *(35)*, reigns supreme in this field, and this is discussed in another chapter of this book (*see* **Chapter 3**).

## 2.3. Protein detection via silver staining

Despite its popularity, Coomassie staining suffers mainly from a relative lack of sensitivity, which is an important issue in proteomics where sample availability is often a problem.

To alleviate this problem, while keeping the ease of use and low costs associated with methods dealing with visible light absorbance, silver staining was introduced a few years after 2D electrophoresis *(36)*. However, it is fair to say that the early days of silver staining were troublesome, with erratic backgrounds and sensitivities, and this was due to the complex chemical mechanisms prevailing in silver staining *(37)*. However, decisive progress was made at the end of the 80's *(38)*, and silver staining is now a reliable technique allying high sensitivity, good reproducibility, low cost and ease of use *(39)*. This is further discussed in another chapter of this book (*see* **Chapter 4**).

The main drawbacks of silver staining in modern proteomics are its limited dynamic range but also its weak compatibility with silver staining *(40, 41)*, although very high performances have been claimed *(42)*. It has been shown that the formaldehyde used in image development was the main culprit *(43)*, and formaldehyde-free protocols have been developed that offer much increased compatibility with mass spectrometry *(44)* (*see* also **Chapter 4**).

**2.3. Protein detection via fluorescence**

To alleviate the problems shown by silver staining and Coomassie Blue, protein detection by fluorescence has been developed and has shown great expansion over the past few years. Opposite to the strict mechanisms that prevail in visible staining, either with Coomassie Blue or with silver, protein detection via fluorescence can been achieved via multiple mechanisms, thereby offering great versatility to this technique.

The first and oldest mechanism is covalent binding, quite often of probes that are not fluorescent but become so when the covalent binding takes place *(45, 46)*. While the performances of such probes were not very impressive, and thus of limited use, a quantum leap was achieved when probes with much higher light absorption and emission characteristics were used. Furthermore, with the development of laser scanners, use of a set of closely related probes could be designed to achieve multiplexing *(47)*, resulting in the very popular DIGE technique *(48)* (*see* also **Chapter 5**).

While this system has shown exquisite performances, it must be kept in mind that only a few fluorescent molecules are bound per protein molecule, resulting in overall low signal intensity for many proteins. This is not a problem for pure detection, as the enormous contrast allows using the full power of laser scanners, but this becomes a problem in some instances, e.g. spot excision for mass spectrometry, where more primitive illumination devices are used, e.g. UV tables.

Thus, another popular mode of protein detection via fluorescence uses non covalent binding, which takes place after migration and therefore does not interfere with protein migration, and which can also take place at many more sites on the proteins. This results in a much higher signal, although the free fluorescent molecule remaining in the gels decreases the contrast compared to covalent binding. Although other candidates have been recently proposed *(49, 50)*, two molecules dominate this field, namely epicocconone *(51)* (e.g. Deep Purple, LavaPurple), and ruthenium based organometallic complexes *(52-54)* (e.g. SYPRO Ruby). These molecules offer a detection sensitivity that is very close to that obtained with silver staining, with a much better linearity and a much better compatibility with mass spectrometry. However, in this latter aspect, they do not perform as well as Coomassie Blue *(41)*.

While these tow modes of detection (covalent and non-covalent binding) are light emission counterparts of modes that have been used in visible detection (light absorption) *(55)*, there is a third detection mode that is specific to fluorescence, which is the use of environment-sensitive probes. These molecules that are used for protein detection do not fluorescence in polar environments such as water, but do fluoresce in less polar environments such as protein-SDS complexes. Several molecules have been shown to achieve protein detection in this general scheme. Protocols using protein fixing and denaturing conditions have been devised with some probes of the styryl class *(56, 57)*, while completely non-denaturing conditions could be used for other probes such as Nile Red *(58)*, carbazolyl vinyl dyes *(59)* and more recently carbocyanines

*(60)*.

While the fixing schemes offer no real advantage over the classical non covalent probes, the nondenaturing schemes offer distinct advantages such as speed, blotting ability *(59, 60)* and more importantly a very good sequence coverage in subsequent mass spectrometry *(60)*.

Last but certainly not least, fluorescence can be used to detect specific motifs on gel-separated proteins, such as sugars *(61)* or phosphate groups *(62)*, thereby offering a very wide palette of detection schemes with wide versatility.

**3. The data analysis issues**

In most instances where 2D gel-based proteomics is used, the production of the gel image by any of the protein detection methods listed above is not the end of the story, and it is very uncommon that all detected protein spots are excised for protein identification by mass spectrometry. Most often, comparative image analysis is used to determine a few spots whose expression profile within the complete experiment matches biologically relevant events. This image analysis process can be split in four majors steps. First data acquisition *(4)*, then spot detection and quantification *(5)*, then gel matching *(6)* (although gel matching can be carried out prior to spot detection in some analysis systems) and finally data analysis *(7)*. It must be stressed that this image analysis process has been used very early after the introduction of 2D gel electrophoresis *(3, 63, 64)*, quite often with sophisticated data analysis tools *(7, 8, 65, 66)*, long before the word proteomics even existed.

Although very cumbersome at these early times, image analysis has dramatically progressed over the years, greatly helped by the considerable increase in computer power. However, image analysis is very dependent on the quality of the experimental data, and especially on their reproducibility. In this respect, the generalization of immobilized pH gradient has had a major impact by bringing a level of positional reproducibility that could never be achieved with conventional isoelectric focusing with ampholytes *(67-69)*. However, image reproducibility is a complex process, and even with the use of immobilized pH gradients, reproducibility is maximized by parallelizing the gels in dedicated instruments *(70)*. Fortunately enough, such parallel electrophoresis instruments had been developed during the early days of 2D electrophoresis *(71, 72)*, when the inter-run variability was very high.

Even though the reproducibility of 2D gel-based proteomics is much higher than for other setups - as shown by higher requirements *(73)* and stricter practices *(74)* - there is an important concern that has emerged over the past few years, i.e. the problem of false positives. Although false positives can have an experimental origin *(75)*, a certain proportion is due to the statistical processes used to determine modulated spots, and thus to the problem of multiple testing *(76-78)*. Although purely statistical tools such as false discovery estimates have been proposed to address this concern *(76-78)*, these tools are not completely well-adapted to the analysis of 2D gel images *(79)*.

In this game of quantitative image analysis to determine spots that show changes in the biological process of interest and thus deserve further studies there is another experimental parameter that plays a key role besides separation reproducibility, namely sample variability and especially biological variability, i.e. from one biological sample to another, before any technical variability introduced by the protein extraction process. This variability grows along with two parameters. One is the plasticity of the proteome, so that variability is often greater in cultured prokaryotic cells than in mammalian ones. The second factor is of course the physiological and genetic heterogeneity, so that in vitro systems are less variable than in vivo systems on inbred laboratory animals, which are in turn less variable than samples obtained in conditions where neither the precise physiological state nor the genetic background can be controlled (typically human samples). In some of the latter cases, the biological variability is so high that it becomes very difficult to find any protein spots showing a statistically significant variation in the experimental process. In such cases, it is tempting to pool samples within the same experimental group, in order to average out the biological variability and facilitate the discovery of modulated proteins. However, it must be kept in mind that interindividual variations are an important part of the problems that do exist in clinics. Thus, this factor cannot be evacuated easily, and pooling must be carefully understood and controlled to achieve correct results *(80, 81)*.

At the beginning of the 21st century, 2D gel-based proteomics is often depicted as an outdated technique, on the basis of its poor ability to analyze membrane proteins *(82)* and of its moderate analysis depth *(83, 84)*. However, in the landscape of proteomic techniques, it still offers unique advantages that make it stand apart (and ahead) of the other proteomic setups for many applications. For example, in the chorus of proteomics setups, 2D gels still offer, and by far, the highest experimental robustness *(74)* and ease of parallelization, mandatory features when large series of samples are to be analyzed, such as in clinical studies or toxicological ones *(85, 86)*. Those points make 2D-gel based applications still useful for many laboratory studies. It is also the only technique that is able to resolve complete proteins with their trail and combination of post-translational modifications, and this has been used in a variety of studies (reviewed in *(21)*). Thus, 2D gel-based proteomics has still a lot to offer to the researchers who will be able to use its strengths.

In conclusion, 2D gel-based proteomics shall not be viewed as a dinosaur in proteomics. In reality in proteomics paleontology it is even much older than the dinosaurs; it is the mammalian reptile (pelycosaurs) of proteomics. And when the giant dinosaurs (shotgun proteomics) will give place to birds (SRM/MRM techniques, *see* **Chapter 16**) for many mainstream proteomic studies, 2D gel-based proteomics will blossom again in its niches, especially every time that its unique capabilities in terms of complete protein separation tool will be useful. With the shrinking of the human genome (barely more complex than the one of the fruit fly) and the recognition of the importance of post-translational modifications in the complexity of the living beings, this age shall come soon.

## 4. Notes

1. General aspects need to be taken into account for reliable protein quantification by 2D electrophoresis:

a) For reliable quantification on-gel protein detection methods must be very sensitive, linear in response, homogeneous and compatible with mass spectrometry.

b) The yield of 2D electrophoresis has been found to be rather moderate (20-40%) depending on the class of protein as well as the efficiency of sample preparation.

2. Several different staining techniques and detection/quantification methods have been developed over the years differing in sensitivity, linearity, homogeneity and compatibility with mass spectrometry. Depending on the scientific question the best method to be applied should be chosen carefully.

3. The reliability and statistical validity of quantification after 2D electrophoresis by automated image analysis depends on the quality and reproducibility of the experimental data. Thus, variation in the whole experimental process from sample preparation to gel generation and data analysis must be kept as small as possible.

4. 2D electrophoresis is the only technique in the bouquet of proteomic methods capable to reproducibly analyze and quantify complete proteins thereby holding the valuable advantage of resolving different protein isoforms.

## 5. References

1. MacGillivray AJ, Rickwood D (1974) The heterogeneity of mouse-chromatin

nonhistone proteins as evidenced by two-dimensional polyacrylamide-gel electrophoresis and ion-exchange chromatography. Eur J Biochem. 41**:**181-190

2. O'Farrell PH (1975) High resolution two-dimensional electrophoresis of proteins. J Biol Chem. 250:4007-4021

3. Anderson NL, Taylor J, Scandora AE et al (1981) The TYCHO system for computer analysis of two-dimensional gel electrophoresis patterns. Clin Chem. 27:1807-1820

4. Vincens P, Paris N, Pujol JL et al (1986) HERMeS: A second generation approach to the automatic analysis of two-dimensional electrophoresis gels Part I: Data acquisition. Electrophoresis 7:347-356

5. Vincens P (1986) HERMeS: A second generation approach to the automatic analysis of two-dimensional electrophoresis gels Part II: Spot detection and integration. Electrophoresis 7:357-367

6. Vincens P, Tarroux P (1987) HERMeS: A second generation approach to the automatic analysis of two-dimensional electrophoresis gels. Part III: Spot list matching. Electrophoresis 8:100-107

7. Tarroux P, Vincens P, Rabilloud T (1987) HERMeS: A second generation approach to the automatic analysis of two-dimensional electrophoresis gels. Part V: Data analysis. Electrophoresis 8:187-199

8. Pun T, Hochstrasser DF, Appel RD et al (1988) Computerized classification of two-dimensional gel electrophoretograms by correspondence analysis and ascendant hierarchical clustering. Appl Theor Electrophor. 1:3-9

9. Matsudaira P (1987) Sequence from picomole quantities of proteins electroblotted onto polyvinylidene difluoride membranes. J Biol Chem. 262:10035-10038

10. Aebersold RH, Leavitt J, Saavedra RA et al (1987) Internal amino acid sequence analysis

of proteins separated by one- or two-dimensional gel electrophoresis after in situ protease digestion on nitrocellulose. Proc Natl Acad Sci U S A. 8:6970-6974

11. Aebersold RH, Pipes G, Hood LE, Ken SBH (1988) N-terminal and internal sequence determination of microgram amounts of proteins separated by isoelectric focusing in immobilized pH gradients. Electrophoresis 9:520-530

12. Rosenfeld J, Capdevielle J, Guillemot JC, Ferrara P (1992) In-gel digestion of proteins for internal sequence analysis after one- or two-dimensional gel electrophoresis. Anal Biochem. 203:173-179

13. Rasmussen HH, van Damme J, Puype M et al (1992) Microsequences of 145 proteins recorded in the two-dimensional gel protein database of normal human epidermal keratinocytes. Electrophoresis 13:960-969

14. Yates JR, Eng JK, McCormack AL, Schieltz D (1995) Method to correlate tandem mass spectra of modified peptides to amino acid sequences in the protein database. Anal Chem. 67:1426-1436

15. Yates JR, McCormack AL, Schieltz D et al (1997) Direct analysis of protein mixtures by tandem mass spectrometry. J Prot Chem. 16:495-497

16. Smolka M, Zhou H, Aebersold R (2002) Quantitative protein profiling using two-dimensional gel electrophoresis, isotope-coded affinity tag labeling, and mass spectrometry. Mol Cell Proteomics. 1:19-29

17. Zhou S, Bailey MJ, Dunn MJ et al (2005) A quantitative investigation into the losses of proteins at different stages of a two-dimensional gel electrophoresis procedure. Proteomics. 5:2739-2747

18. Santoni V, Molloy M, Rabilloud T (2000) Membrane proteins and proteomics: un amour impossible? Electrophoresis 21:1054-1070

19. Eravci M, Fuxius S, Broedel O et al (2008) The whereabouts of transmembrane proteins from rat brain synaptosomes during two-dimensional gel electrophoresis. Proteomics 8: 1762-1770

20. Rabilloud T, Vaezzadeh AR, Potier N et al (2009) Power and limitations of electrophoretic separations in proteomics strategies. Mass Spectrom Rev. 28:816-843

21. Rabilloud T, Chevallet M, Luche S, Lelong C (2010) Two-dimensional gel electrophoresis in proteomics: Past, present and future. J Proteomics. 73:2064-2077

22. Rabilloud T (2010) Variations on a theme: changes to electrophoretic separations that can make a difference. J Proteomics. 73:1562-1572

23. Miller I, Crawford J, Gianazza E (2006) Protein stains for proteomic applications: which, when, why? Proteomics 6:5385-5408

24. Bonner WM, Laskey RA (1974) A film detection method for tritium-labelled proteins and nucleic acids in polyacrylamide gels. Eur J Biochem. 46:83-88

25. Perng GS, Rulli RD, Wilson DL, Perry GW (1988) A comparison of fluorographic methods for the detection of 35S-labeled proteins in polyacrylamide gels. Anal Biochem. 173:387-392

26. Laskey RA, Mills AD (1975) Quantitative film detection of 3H and 14C in polyacrylamide gels by fluorography. Eur J Biochem. 56:335-341

27. Duncan R, McConkey EH (1982) How many proteins are there in a typical mammalian cell? Clin Chem. 28:749-755

28. Bravo R, Fey SJ, Bellatin J et al (1981) Identification of a nuclear and of a cytoplasmic polypeptide whose relative proportions are sensitive to changes in the rate of cell proliferation. Exp Cell Res. 136:311-319

29. Bravo R, Frank R, Blundell PA, Macdonald-Bravo H (1987) Cyclin/PCNA is the

auxiliary protein of DNA polymerase-delta. Nature 326:515-517

30. Bravo R, Fey SJ, Small JV et al (1981) Coexistence of three major isoactins in a single sarcoma 180 cell. Cell 25:195-202

31. Sobel A, Tashjian AH Jr. (1983) Distinct patterns of cytoplasmic protein phosphorylation related to regulation of synthesis and release of prolactin by GH cells. J Biol Chem. 258:10312-10324

32. Patterson SD, Latter GI (1993) Evaluation of storage phosphor imaging for quantitative analysis of 2-D gels using the Quest II system. Biotechniques 15:1076-1083

33. Zhou SB, Mann CJ, Dunn MJ et al (2006) Measurement of specific radioactivity in proteins separated by two-dimensional gel electrophoresis. Electrophoresis 27:1147-1153

34. Choi JK, Tak KH, Jin LT et al (2002) Background-free, fast protein staining in sodium dodecyl sulfate polyacrylamide gel using counterion dyes, zincon and ethyl violet. Electrophoresis 23:4053-4059

35. Neuhoff V, Arold N, Taube D, Ehrhardt W (1988) Improved staining of proteins in polyacrylamide gels including isoelectric focusing gels with clear background at nanogram sensitivity using Coomassie Brilliant Blue G-250 and R-250. Electrophoresis. 9:255-262

36. Switzer RC, Merril CR, Shifrin S (1979) A highly sensitive silver stain for detecting proteins and peptides in polyacrylamide gels. Anal Biochem. 98:231-237

37. Rabilloud T (1990) Mechanisms of protein silver staining in polyacrylamide gels: a 10-year synthesis. Electrophoresis 11:785-794

38. Blum H, Beier H, Gross H (1987) Improved silver staining of plant proteins, RNA and DNA in polyacrylamide gels. Electrophoresis 8:93-99

39. Chevallet M, Luche S, Rabilloud T (2006) Silver staining of proteins in polyacrylamide

gels. Nat Protoc. 1:1852-1858

40. Scheler C, Lamer S, Pan Z et al (1998) Peptide mass fingerprint sequence coverage from differently stained proteins on two-dimensional electrophoresis patterns by matrix assisted laser desorption/ionization-mass spectrometry (MALDI-MS). Electrophoresis 19:918-927

41. Chevalier F, Centeno D, Rofidal V et al (2006) Different impact of staining procedures using visible stains and fluorescent dyes for large-scale investigation of proteomes by MALDI-TOF mass spectrometry. J Proteome Res. 5:512-520

42. Shevchenko A, Wilm M, Vorm O, Mann M (1996) Mass spectrometric sequencing of proteins silver-stained polyacrylamide gels. Anal Chem. 68:850-858

43. Richert S Luche S Chevallet M et al (2004) About the mechanism of interference of silver staining with peptide mass spectrometry. Proteomics 4:909-916

44. Chevallet M, Luche, S, Diemer H et al (2008) Sweet silver: a formaldehyde-free silver staining using aldoses as developing agents, with enhanced compatibility with mass spectrometry. Proteomics 8:4853-4861

45. Jackowski G, Liew CC (1980) Fluorescamine staining of nonhistone chromatin proteins as revealed by two-dimensional polyacrylamide gel electrophoresis. Anal Biochem. 102:321-325

46. Urwin VE, Jackson P (1991) A multiple high-resolution mini two-dimensional polyacrylamide gel electrophoresis system: imaging two-dimensional gels using a cooled charge-coupled device after staining with silver or labeling with fluorophore. Anal Biochem. 195:30-37

47. Unlu ., Morgan ME, Minden JS (1997) Difference gel electrophoresis: a single gel method for detecting changes in protein extracts. Electrophoresis 18:2071-2077

48. Tonge R, Shaw J, Middleton B et al (2001) Validation and development of fluorescence

two-dimensional differential gel electrophoresis proteomics technology. Proteomics 1:377-396

49. Suzuki Y, Yokoyama K (2008) Design and synthesis of ICT-based fluorescent probe for high-sensitivity protein detection and application to rapid protein staining for SDS-PAGE. Proteomics 8:2785-2790

50. Cong WT, Jin LT, Hwang SY, Choi JK (2008) Fast fluorescent staining of protein in sodium dodecyl sulfate polyacrylamide gels by palmatine. Electrophoresis 29:417-423

51. Mackintosh JA, Choi HY, Bae SH et al (2003) A fluorescent natural product for ultra sensitive detection of proteins in one-dimensional and two-dimensional gel electrophoresis. Proteomics 3:2273-2288

52. Berggren K, Chernokalskaya E, Steinberg TH et al (2000) Background-free, high sensitivity staining of proteins in one- and two-dimensional sodium dodecyl sulfate-polyacrylamide gels using a luminescent ruthenium complex. Electrophoresis 21:2509-2521

53. Rabilloud T, Strub JM, Luche S et al (2001) A comparison between Sypro Ruby and ruthenium II tris (bathophenanthroline disulfonate) as fluorescent stains for protein detection in gels. Proteomics 1:699-704

54. Lamanda A, Zahn A, Roder D, Langen H (2004) Improved Ruthenium II tris (bathophenantroline disulfonate) staining and destaining protocol for a better signal-to-background ratio and improved baseline resolution. Proteomics 4:599-608

55. Fazekas SDS, Webster RG, Datyner A (1963) Two new staining procedures for quantitative estimation of proteins on electrophoretic strips. Biochim Biophys Acta 71:377-39.

56. Steinberg TH, Jones LJ, Haugland RP, Singer VL (1996) SYPRO orange and SYPRO red

protein gel stains: one-step fluorescent staining of denaturing gels for detection of nanogram levels of protein. Anal Biochem. 239:223-237

57. Malone JP, Radabaugh MR, Leimgruber RM, Gerstenecker GS (2001) Practical aspects of fluorescent staining for proteomic applications. Electrophoresis 22:919-932

58. Daban JR, Bartolome S, Samso M (1991) Use of the hydrophobic probe Nile red for the fluorescent staining of protein bands in sodium dodecyl sulfate-polyacrylamide gels. Anal Biochem. 199:169-174

59. Steinberg TH, Lauber WM, Berggren K et al (2000) Fluorescence detection of proteins in sodium dodecyl sulfate-polyacrylamide gels using environmentally benign, nonfixative, saline solution. Electrophoresis 21:497-508

60. Luche S, Lelong C Diemer H et al (2007) Ultrafast coelectrophoretic fluorescent staining of proteins with carbocyanines. Proteomics 7:3234-3244

61. Hart C, Schulenberg B, Steinberg TH et al (2003) Detection of glycoproteins in polyacrylamide gels and on electroblots using Pro-Q Emerald 488 dye, a fluorescent periodate Schiff-base stain. Electrophoresis 24:588-598

62. Schulenberg B, Goodman TN, Aggeler R et al (2004) Characterization of dynamic and steady-state protein phosphorylation using a fluorescent phosphoprotein gel stain and mass spectrometry. Electrophoresis 25:2526-2532

63. Garrels JI (1979) Two dimensional gel electrophoresis and computer analysis of proteins synthesized by clonal cell lines. J Biol Chem. 254:7961-7977

64. Vo KP, Miller MJ, Geiduschek EP et al (1981) Computer analysis of two-dimensional gels. Anal Biochem. 112:258-271

65. Tarroux P (1983) Analysis of protein patterns during differentiation using 2-D electrophoresis and computer multidimensional classification. Electrophoresis 4:63-70

66. Appel R, Hochstrasser D, Roch C et al (1988) Automatic classification of two-dimensional gel electrophoresis pictures by heuristic clustering analysis: a step toward machine learning. Electrophoresis 9:136-142

67. Corbett JM, Dunn MJ, Posch A, Gorg A (1994) Positional reproducibility of protein spots in two-dimensional polyacrylamide gel electrophoresis using immobilised pH gradient isoelectric focusing in the first dimension: an interlaboratory comparison. Electrophoresis 15:1205-1211

68. Blomberg A, Blomberg L, Norbeck J et al (1995) Interlaboratory reproducibility of yeast protein patterns analyzed by immobilized pH gradient two-dimensional gel electrophoresis. Electrophoresis 16:1935-1945

69. Choe LH, Lee KH (2003) Quantitative and qualitative measure of intralaboratory two-dimensional protein gel reproducibility and the effects of sample preparation, sample load, and image analysis. Electrophoresis 24:3500-3507

70. Eravci M, Fuxius S, Broedel O et al (2007) Improved comparative proteome analysis based on two-dimensional gel electrophoresis. Proteomics 7:513-523

71. Anderson NG, Anderson NL (1978) Analytical techniques for cell fractions. XXI. Two-dimensional analysis of serum and tissue proteins: multiple isoelectric focusing. Anal Biochem. 85:331-340

72. Anderson NL, Anderson NG (1978) Analytical techniques for cell fractions. XXII. Two-dimensional analysis of serum and tissue proteins: multiple gradient-slab gel electrophoresis. Anal Biochem. 85:341-354

73. Celis JE (2004) Gel-based proteomics: what does MCP expect? Mol Cell Proteomics 3:949

74. Hackett M (2008) Science, marketing and wishful thinking in quantitative proteomics.

Proteomics 8:4618-4623

75. Fuxius S, Eravci M, Broedel O et al (2008) Technical strategies to reduce the amount of "false significant" results in quantitative proteomics. Proteomics 8:1780-1784

76. Karp NA, Lilley KS(2007) Design and analysis issues in quantitative proteomics studies. Proteomics 7:42-50

77. Karp NA, McCormick PS, Russell MR, Lilley KS (2007) Experimental and statistical considerations to avoid false conclusions in proteomics studies using differential in-gel electrophoresis. Mol Cell Proteomics 6:1354-1364

78. Eravci M, Mansmann U, Broedel O et al (2009) Strategies for a reliable biostatistical analysis of differentially expressed spots from two-dimensional electrophoresis gels. J Prot Res. 8:2601-2607

79. Diz AP, Carvajal-Rodriguez A, Skibinski DO (2011) Multiple hypothesis testing in proteomics: a strategy for experimental work. Mol Cell Proteomics 10:M110.004374

80. Diz AP, Truebano M, Skibinski DO (2009) The consequences of sample pooling in proteomics: an empirical study. Electrophoresis 30:2967-2975

81. Karp NA, Lilley KS (2009) Investigating sample pooling strategies for DIGE experiments to address biological variability. Proteomics 9:388-397

82. Rabilloud T (2009) Membrane proteins and proteomics: love is possible, but so difficult. Electrophoresis 30:S174-180

83. Petrak J, Ivanek R, Toman O et al (2008) Déjà vu in proteomics. A hit parade of repeatedly identified differentially expressed proteins. Proteomics 8:1744-1749

84. Wang P, Bouwman FG, Mariman EC (2009) Generally detected proteins in comparative proteomics--a matter of cellular stress response? Proteomics 9:2955-2966

85. Aicher L, Wahl D, Arce A et al (1998) New insights into cyclosporine A nephrotoxicity

by proteome analysis. Electrophoresis 19:1998-2003

86. Anderson NL, EsquerBlasco R, Richardson F et al (1996) The effects of peroxisome proliferators on protein abundances in mouse liver. Toxicol Appl Pharmacol. 137:75-89